\documentstyle{pss}
\input epsf.tex
\begin{document}

\title{ Jahn-Teller Effect and Superexchange \\
        in Half-Doped Manganites }

\author{{\sc Olga Sikora}\footnote{) E-mail:
sikora@alphetna.if.uj.edu.pl}\hskip .4cm  and {\sc Andrzej M. Ole\'s}}

\address{ Marian Smoluchowski Institute of Physics, Jagellonian 
          University,\\
          Reymonta 4, PL-30059 Krak\'ow, Poland}

\submitted{1 July 2002} 
\maketitle
\hspace{9mm} 
Subject classification: 71.20.-b, 75.47.Gk, 75.50.Ee

\begin{abstract}
We investigate the stability of the charge exchange (CE) phase within 
a microscopic model which describes a single plane as in 
La$_{0.5}$Sr$_{1.5}$MnO$_4$. The model includes Coulomb interactions 
(on-site and intersite), the Jahn-Teller term and the superexchange 
interactions due to $e_g$ and $t_{2g}$ electrons. By investigating the 
phase diagram at $T=0$ in mean-field approximation we conclude that 
the superexchange interactions can stabilize the CE phase, but only 
if they are stronger than estimated from spectroscopy.\\
Journal reference: O. Sikora and A. M. Ole\'s, 
                   Phys. Stat. Sol. (b) {\bf 236}, 380 (2003). 
\end{abstract}

The doped manganites belong to a very interesting class of transition 
metal oxides, with orbital degrees of freedom and several magnetic 
phases stable in various doping regimes. The so-called charge exchange 
(CE) phase, composed of one-dimensional (1D) ferromagnetic (FM) zigzag chains 
with an antiferromagnetic (AF) coupling between them, has attracted a 
lot of attension recently, and the origin of its stability is still 
under debate \cite{Pop02}. The charge and orbital ordering sets in 
La$_{0.5}$Ca$_{0.5}$MnO$_3$ at 
$T_{\rm CO}\simeq 225$ K, and is followed by a magnetic transition at 
$T_N\simeq 155$ K \cite{Rad97}. The CE-phase was also observed in 
one-plane La$_{0.5}$Sr$_{1.5}$MnO$_4$ compound, with a similar sequence 
of phase transitions (a structural transition at $T_{\rm CO}=255$ K, 
a magnetic transition at $T_N\simeq 110$ K \cite{Ste96}), and a rather 
pronounced orbital order \cite{Mur98}. Recent experiments show that the 
charge order is particularly pronounced in this case, and the Jahn-Teller 
(JT) distortions around the occupied Mn$^{3+}$ centers are induced 
\cite{Mah01}. The reasons for appearing of the CE phase in 
La$_{0.5}$Sr$_{1.5}$MnO$_4$ are not yet fully understood.

\begin{figure}[h]
\epsfxsize=13cm
\epsfbox{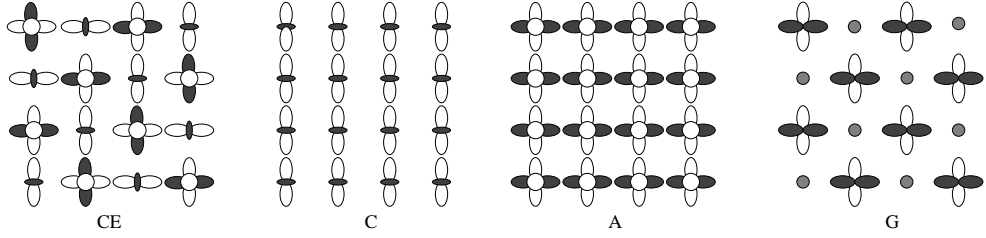}
\caption{Orbital ordering in considered magnetic phases of half-doped
manganites: CE, C, A, and G phase.} 
\label{test3}
\end{figure}

We investigate the stability of the single-plane CE-type phase with 
respect to three other phases: the C phase with staggered linear FM 
chains, the $G$-type AF (G) phase, and the FM plane of the $A$-type AF 
(A) phase. These phases are characterized by different orbital structure,
as shown in Fig. \ref{test3}. The tight binding model without any 
interactions applied to the CE phase gives a band insulator due to 
a particular conflict of orbital phases which {\it frustrates\/} the 
kinetic energy \cite{Bri99}. This favors the observed zigzag CE phase 
with respect to the C phase. The CE phase is destabilized by large 
on-site Coulomb interactions $U$, and we have shown that either Coulomb 
intersite or the JT interaction can stabilize the CE phase \cite{OS02}. 
This effect is enhanced when the superexchange energy is included. 
However, when we take into account $A$-AF phase or $G$-AF phase, the 
energy of CE phase is higher for $U$ not bigger then about $7t$ and 
realistic values of other parameters. We have verified that the CE phase 
can be stabilized in the present model only by increasing the 
superexchange above that which follows from the spectroscopic parameters.  

We consider below a model which describes the electronic properties of 
the single-plane compounds (La$_{0.5}$Sr$_{1.5}$MnO$_4$), and 
includes spin, orbital and charge degrees of freedom, as well as the 
cooperative JT distortions. We consider the Hamiltonian composed of 
four terms: 
\begin{equation}
\label{model} 
{\cal H} =  H_{t} + H_{int} + H_{\rm JT} + H_{\rm SE},   
\end{equation}
where $H_{t}$ refers to kinetic energy of $e_g$ electrons hopping along
FM bonds (the AF bonds cannot contribute as the kinetic energy is blocked 
by double exchange), $H_{int}$ describes the Coulomb interactions, 
$H_{\rm JT}$ -- the JT energy, and $H_{\rm SE}$ -- the superexchange 
energy. The superexchange follows from $d_i^n d_j^m\rightleftharpoons 
d_i^{n+1}d_j^{m-1}$ charge excitations for the Mn$^{3+}$-Mn$^{3+}$, 
Mn$^{3+}$-Mn$^{4+}$, and Mn$^{4+}$-Mn$^{4+}$ pairs \cite{Amo02}.

The hopping Hamiltonian for CE phase is given by \cite{Bri99}:
\begin{equation}
\label{Ht} 
H_t = -\sum_{i\in B,j\in C}\Big[ (-1)^{\lambda_{ij}}
t_1b_i^{\dagger}a_{jx}^{}+t_2b_i^{\dagger}a_{jz}^{}+{\rm H.c.}\Big], 
\end{equation}    
where $B$ and $C$ refer to the bridge and corner positions along the 1D 
zigzag chain, respectively, and $t_1=\frac{\sqrt{3}}{2}t$, 
$t_2=\frac{1}{2}t$ are the hopping elements, with $t$ standing for the 
hopping between two directional orbitals along the bond (e.g. two 
$3x^2-r^2$ orbitals for $\langle ij\rangle\parallel a$) and being the 
energy unit ($t\simeq 0.41$ eV was obtained from the charge-transfer 
model \cite{Fei99}). The phase factor $(-1)^{\lambda_{ij}}$ follows 
from the orbital phases. The minimal basis set includes a single orbital 
($3x^2-r^2$ or $3y^2-r^2$) at bridge positions ($b_i^{\dagger}$), and 
two orbitals: 
$|x\rangle\equiv|x^2-y^2\rangle$ and $|z\rangle\equiv|3z^2-r^2\rangle$ 
at corner sites ($a_{jx}^{\dagger}$ and $a_{jz}^{\dagger}$), considered 
in the tight-binding model of Solovyev \cite{Sol01}.

In our calculations we include the on-site Coulomb interaction 
$\propto U$, and either the intersite Coulomb interaction $\propto V$, 
or the energy of JT distortions $\propto E_{\rm JT}$:
\begin{eqnarray}
\label{Hu}
H_{int}&=&U\sum_{j\in C}n_{jx}n_{jz}+V\sum_{\langle ij\rangle}n_{i}n_{j},\\
\label{Hjt} 
H_{\rm JT}&=&E_{\rm JT}\Big[\sum_{i,j(i)}q_i(n_{i\zeta}-n_{i\xi})(1-n_j)
             +\textstyle{\frac{1}{2}}\sum_{i}q_i^2\Big].
\end{eqnarray}
We treat the JT interactions as lattice polarons and thus introduce the 
local variables $\{q_i\}$ which induce the charge-hole correlations 
$\propto n_i(1-n_j)$. They split the $e_g$ orbitals around a hole, with 
the directional orbital $|\zeta\rangle$ oriented along the bond being 
favored with respect to the planar orbital $|\xi\rangle$ \cite{Kil99}. 
For instance, for a bond $\langle ij\rangle\parallel c$ these orbitals 
are: $3z^2-r^2$ and $x^2-y^2$.   

\begin{figure}[h]
\epsfxsize=12cm
\epsfbox{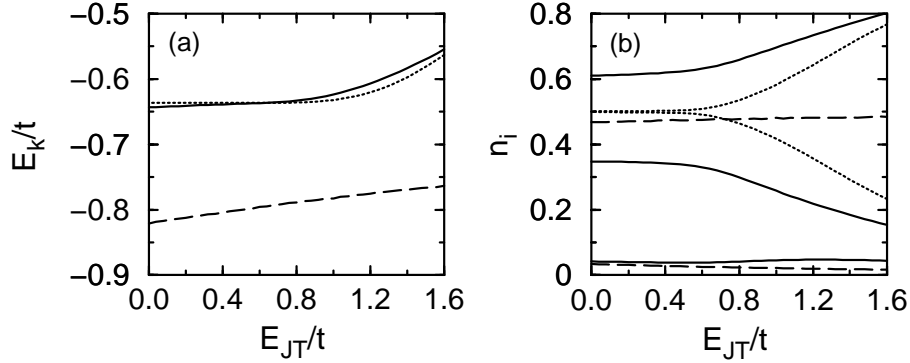}
\caption{Kinetic energy per site (a) and electron densities (b), as 
functions of the JT interaction at $U=6t$; solid, dashed and dotted 
lines refer to the CE, A and C phase. In (b) the electron densitied 
$n_i$ are plotted (from the top) for bridge, $|x\rangle$ and $|z\rangle$ 
orbital (CE phase); $|x\rangle$ and $|z\rangle$ orbitals (A phase), 
and two neighboring orbitals in the C phase. For G phase the kinetic 
energy is equal to zero and the orbital occupation is constant.}
\label{ekior}
\end{figure}

We have solved the model (\ref{model}) in the mean-field approximation 
(MFA). Fig. \ref{ekior} shows electron densities and kinetic energy per 
site obtained for $U=6t$ and for different values of $E_{\rm JT}$. For 
the CE phase, the on-site Coulomb interaction $U$ induces small charge 
ordering, which becomes much stronger when the JT (or nearest-neighbor 
Coulomb) interaction is included. In the C phase the charge distribution 
is symmetric for small values of interaction constants, while when the
interactions increase, the electrons gradually localize. The absolute 
value of the kinetic energy of these two phases becomes lower with 
increasing $V$ or $E_{\rm JT}$. The kinetic energy of the A phase is 
much lower for all values of the JT interaction constant $E_{\rm JT}$, 
and we conclude that neither Coulomb nor JT interactions can stabilize 
the CE phase.  

In Table 1 we show the superexchange energy contributions for 
different phases. For anisotropic phases we show the values obtained 
for FM and AF bonds. As one can expect, the SE energy of the G phase 
has the biggest absolute value, and the SE contribution to the energy 
of the FM A phase is the smallest. Due to the orbital ordering (Fig. 
\ref{test3}), the SE interactions favor the CE phase to the C phase 
(both phases have two AF and two FM bonds). Therefore, as long as the 
JT energy does not dominate in the G phase due to its orbital pattern 
(Fig. \ref{test3}), the superexchange can stabilize the CE phase.    
 
\begin{table}[h]
\label{SE}
\caption{Superexchange energy per one site in units of 10$^{-3}t$ at 
$U=6t$ and $E_{\rm JT}=1.2t$, as obtained using the spectroscopic 
parameters given in Ref. \protect\cite{Fei99}. For the CE phase two 
anisotropic contributions: FM along the zigzag chain ($\parallel$) and 
AF between two neighboring chains ($\perp$) are given separately. The 
FM bonds in the C phase do not contribute any superexchange energy.}
\begin{center} \begin{tabular}{|c|c|c|c|c|c|c|}
\hline
& & & & & &\\
excitation & ion pair & CE & CE & C & G & A \\
& & $ \parallel $ & $\perp $ & $\perp $ & & 
\\ \hline
$e_g$    & Mn$^{3+}$-Mn$^{3+}$ & -7.1 & -10.4 &-6.9  & 0     & -21.1\\
$e_g$    & Mn$^{3+}$-Mn$^{4+}$ &    0 & -17.5 &-13.1 & -72.4 & 0\\
$t_{2g}$ & Mn$^{3+}$-Mn$^{3+}$ & 0 & -2.0  &-2.4  & 0     & 0\\
$t_{2g}$ & Mn$^{4+}$-Mn$^{4+}$ & 0 & -4.4  &-5.2  & 0     & 0\\
$t_{2g}$ & Mn$^{3+}$-Mn$^{4+}$ & 0 & -16.4 &-14.5 & -53.7 & 0\\
\hline
& total: &-7.1 & -50.7 & -42.1 & -126.1 & -21.1 \\ 
\hline
\end{tabular} \end{center}
\end{table}

\begin{figure}[h]
\epsfxsize=12cm
\epsfbox{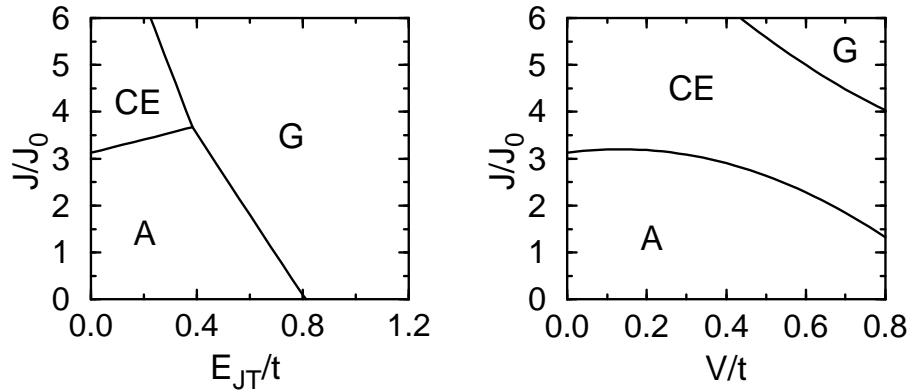}
\caption{Phase diagrams obtained in MFA with $U=6t$ by including either 
the JT interaction $E_{\rm JT}$ (left), or the intersite Coulomb 
interaction $V$ (right).}
\label{fazy}
\end{figure}

We compared the energies obtained for four phases of Fig. \ref{test3} as 
functions of either the JT interaction $E_{\rm JT}$, or the intersite 
Coulomb interaction $V$, including the superexchange contributions scaled 
by a multiplicative factor $J/J_0$, with $J_0=t^2/U\simeq 23$ meV
\cite{Fei99}. We have shown that if only the C and the CE phase is 
considered, for different values of $U$ the JT interactions cause very 
similar effects as the intersite Coulomb interaction with $V\approx 0.5 
E_{\rm JT}$ \cite{OS02}. Fig. \ref{fazy} shows phase diagrams including 
also the A and the G phase and they are significantly different. For low 
values of interactions and without the superexchange, the A phase is 
stabilized by the kinetic energy. Increasing values of the $E_{\rm JT}$ 
favor the G phase and it is the most stable phase above $E_{\rm JT}\approx 
0.7t$ for $J=J_0$. The superexchange interactions can stabilize the CE 
phase only for small values of JT interaction constant $E_{\rm JT}<0.4t$. 
In contrast, the intersite Coulomb interaction $V$ is not sensitive to the 
type of orbital occupied and the G phase is here less stable. It becomes 
the most stable phase only for very strong Coulomb interactions, but the 
range of stability range of the CE phase is much bigger then in the former 
case. We note that the sequence of phases: A, CE and G, obtained at 
increasing $J/J_0$ is the same as found by Dagotto {\it et al.} 
\cite{Hot00}, showing that this result is robust and does not depend on 
the accurate form of $H_{\rm SE}$. However, the C phase is never stable 
in our model, contrary to the model which includes the non-cooperative JT 
effect instead \cite{Hot00}.  
 
We conclude that the superexchange interactions play an important role in 
stabilizing the CE phase, but full understanding of microscopic reasons 
of its stability needs further investigation. It is expected that the 
CE phase could be more stable by going beyond the MFA and including the 
correlation effects for larger values of $U$.

This work was financially supported by KBN Project No. 5~P03B~055~20.

\end{document}